\documentclass[twocolumn,aps,showpacs]{revtex4}
\usepackage{amssymb}
\usepackage{amsmath,bm}
\usepackage{graphicx}
\usepackage[normalem]{ulem}
\usepackage{multirow}

\usepackage[usenames, dvipsnames]{xcolor}
\usepackage{tensor}

\renewcommand{\sout}{\bgroup \color{red} \ULdepth=-.5ex \ULset}

\newcommand{\pdt}{$ N_\text{t}N_\text{p}/ N_\text{d}^2~$}

\begin{document}

\title{Enhanced yield ratio of light nuclei in heavy ion collisions with a first-order QCD phase transition}
\author{Kai-Jia Sun\footnote{%
kjsun@tamu.edu}}
\affiliation{Cyclotron Institute and Department of Physics and Astronomy, Texas A\&M University, College Station, Texas 77843, USA}
\author{Feng Li\footnote{%
fengli@lzu.edu.cn}}
\affiliation{School of Physical Science and Technology, Lanzhou University, Lanzhou, Gansu, 073000, China }
\author{Jun Xu\footnote{%
 xujun@zjlab.org.cn}}
 \affiliation{Shanghai Advanced Research Institute, Chinese Academy of Sciences, Shanghai 201210, China}
\affiliation{Shanghai Institute of Applied Physics, Chinese Academy of Sciences, Shanghai 201800, China}
\author{Che Ming Ko\footnote{%
ko@comp.tamu.edu}}
\affiliation{Cyclotron Institute and Department of Physics and Astronomy, Texas A\&M University, College Station, Texas 77843, USA}
\author{Lie-Wen Chen\footnote{%
lwchen@sjtu.edu.cn}}
\affiliation{School of Physics and Astronomy and Shanghai Key Laboratory for Particle Physics and Cosmology,Shanghai Jiao Tong University,  Shanghai  200240,  China}

\date{\today}

\begin{abstract}
Using a transport model that includes a first-order chiral phase transition between the partonic and the
hadronic matter, we study the development of density fluctuations in the matter produced  in heavy ion collisions  as it  undergoes the phase transition,  and their time evolution in later hadronic stage of the collisions.   Using the coalescence model to describe  the production of deuterons and tritons from
nucleons at the kinetic freeze out, we find that the yield ratio $ N_\text{t}N_\text{p}/ N_\text{d}^2$, where
$N_\text{p}$, $N_\text{d}$, and $N_\text{t}$ are, respectively, the proton, deuteron, and triton numbers,  is enhanced if the evolution trajectory of the  produced matter in the QCD phase diagram passes through the spinodal region of a first-order chiral phase transition.
\end{abstract}

\pacs{12.38.Mh, 5.75.Ld, 25.75.-q, 24.10.Lx}
\maketitle

\section{Introduction}
\label{introduction}

It is generally agreed that the   quark-gluon plasma (QGP), which is believed to have existed at the very beginning of our universe, has been created in collisions of two heavy nuclei at extremely high energies~\cite{Shu80,BraunMunzinger:2008tz,Braun-Munzinger:2015hba,Yin:2018ejt,Bzdak:2019pkr,Luo:2020pef}. In these collisions,  the produced QGP undergoes a rapid expansion, leading to a decrease of its temperature until it is converted to a hadronic matter (HM) as a result of   color confinement. The hadronic matter further expands and cools due to scatterings among hadrons until the kinetic freeze out when  they suffer their last scatterings.   According to calculations  based on the lattice quantum chromodynamics (LQCD) at zero and small baryon chemical potentials~\cite{Bernard:2004je,Aok06,Baz12}, the phase transition from the QGP to HM is a smooth crossover.  At finite baryon chemical potential ($\mu_B$) or baryon density ($\rho_B$), many effective theories have suggested, however, that the quark-hadron phase transition  is a first-order one~\cite{Asa89,Fuk08,Car10,Bra12,Stephanov:2004wx,Stephanov:2007fk,Fukushima:2013rx,Baym:2017whm}, indicating the existence of a critical endpoint (CEP) on  the first-order phase transition line in  the $\mu_B-T$ plane. By changing  the  beam energy as well as selecting the size of the collision system and the  rapidity range of produced particles,  heavy ion collisions provide the possibility of exploring different regions in the $\mu_B-T$ (or $\rho_B-T$) plane of the QCD phase diagram. To search for the CEP and locate the phase boundary in the QCD phase diagram are the main motivations for the  experiments being carried out in the Beam Energy Scan (BES) program at RHIC, FAIR, NICA, and NA61/SHINE.

In heavy-ion collisions, the created matter is expected to develop  large  baryon density fluctuations as a result of the spinodal instability~\cite{Mishustin:1998eq,Chomaz:2003dz,Sasaki:2007db,Steinheimer:2012gc,Herold:2013qda,Li:2016uvu,Nahrgang:2016eou,Steinheimer:2019iso} or the long-range correlations~\cite{Stephanov:1998dy,Stephanov:2011pb} if its evolution trajectory in the QCD phase diagram passes across  the first-order phase transition line or   the critical region of CEP. Although studying density fluctuations in heavy ion collisions offers the opportunity to locate the CEP in the QCD phase diagram, it is  a challenge
to identify observables that are sensitive to these density fluctuations or correlations and thus allow us to extract information on the phase structure of QCD from experimental measurements~\cite{Berdnikov:1999ph,Palhares:2010zz,Ohnishi:2016bdf,Mukherjee:2016kyu,Yin:2018ejt,Bluhm:2020mpc,Asakawa:2019kek,Bzdak:2019pkr,Luo:2020pef}.   For the recently developed technique based on the machine learning with deep neural network, it provides a plausible way to discriminate results from different equations of state with no  pre-defined physical observables~\cite{Pang:2016vdc,Du:2019civ}  but has not yet been successfully applied to real experimental data. On the other hand, a well-known observable is the fourth-order fluctuation $\kappa\sigma^2$ in the net-proton multiplicity distribution, given by the product of its kurtosis $\kappa$ and the variance $\sigma^2$, which is expected to show a non-monotonic behavior in its collision energy dependence~\cite{Stephanov:2008qz,Stephanov:2011pb}.  Another promising observable is the  production of light nuclei, such as  deuteron~($d$), triton ($t$), helium-3 ($^3$He) etc., due to their composite structures~\cite{Steinheimer:2012gc,Steinheimer:2013xxa,Sun:2017xrx,Sun:2018jhg,Shuryak:2018lgd,Shuryak:2019ikv}.  In particular, it was shown in Refs.~\cite{Sun:2017xrx,Sun:2018jhg} that the  ratio $\mathcal{O}_\text{p-d-t} = N_\text{t}N_\text{p}/ N_\text{d}^2$ of the proton, deuteron, and triton yields would be enhanced and a peak structure  is expected in its collision energy dependence if large spatial density fluctuations are developed during the QGP to HM phase transition and can survive later hadronic scatterings.  Recently, this ratio has been measured  in heavy ion collisions  at  both  SPS energies~\cite{Anticic:2016ckv,Sun:2017xrx} and RHIC BES energies~\cite{Adam:2019wnb,Zhang:2019wun,Zhang:2020ewj}, and preliminary data indeed shows possible non-monotonic peak structures in   its collision energy dependence.  Although it has been shown in Ref.~\cite{Li:2016uvu}  that large spatial density fluctuations can develop during the QGP to HM phase transition, whether they can survive later hadronic scatterings has not been studied. To answer the question of whether the non-mononic behavior of \pdt seen in experimental data is due to a first-order or second-order QGP to HM phase transition thus requires detailed dynamical modelings of the time evolution of  density fluctuations in heavy ion collisions~\cite{Nahrgang:2015tva,Bluhm:2020mpc}.

Various  hydrodynamic approaches are being developed to include the effect of density fluctuations, and they include   the stochastic fluid dynamics~\cite{Murase:2016rhl,Hirano:2018diu,Nahrgang:2017oqp,Bluhm:2018plm,Singh:2018dpk},  the hydrokinetic approach~\cite{Akamatsu:2016llw,An:2019osr,Stephanov:2017ghc,Rajagopal:2019xwg}, and   the nonequilibrium chiral fluid dynamics~\cite{Nahrgang:2011mg,Herold:2013bi,Herold:2013qda,Nahrgang:2016eou,Herold:2014zoa,Nahrgang:2018afz}.  Instead of the hydrodynamic model, we  extend in the present study the partonic transport  model of Refs.~\cite{Li:2016uvu,Xu:2013sta}, which includes the dynamics of baryon density fluctuations in  the partonic
phase, to include also the effect of hadronic scatterings on the baryon density fluctuations.  The partonic phase described in Refs.~\cite{Li:2016uvu,Xu:2013sta} is based on the
Nambu-Jona-Lasino (NJL) model~\cite{Nam611,Nam612}, which consists of scalar and vector interactions  between light quarks in the mean-field  approximation. Without the vector interaction, the chiral phase transition  in this model is a smooth  crossover at small baryon chemical potentials but changes to a first-order  transition as the baryon chemical potential becomes large.  Because of the repulsive nature of quark vector interaction in baryon-rich quark matter, increasing its strength  can change the location of the critical endpoint in the phase diagram of quark matter and even make it disappear. With proper initializations of the baryon-rich quark matter produced in heavy ion collisions, its evolution trajectory  for the case of a first-order phase transition in its equation of state could enter the spinodal region of its phase diagram and develops large density fluctuations~\cite{Li:2016uvu}. Converting these quarks to hadrons via the  spatial  coalescence model in a multiphase transport (AMPT) model~\cite{Lin05}, the quark density fluctuations are transferred to the density fluctuations    in the   initially produced   hadronic matter. The effects of subsequent hadronic scatterings can then be modeled by a relativistic transport (ART) model~\cite{Li:1995pra} in the AMPT to study if the baryon density fluctuations  can survive during the expansion of the hadronic matter.

Based on the nucleon coalescence model using the kinetically freeze-out nucleons from the above described approach for light nuclei production,  we study in this work the effects of  baryon density fluctuations induced during the first-order chiral phase transition of quark matter on their production  in heavy ion collisions. We find that these density fluctuations  can survive a  long time and eventually leads to a yield ratio of \pdt of around 0.5  that is larger than those predicted in  a  recent study~\cite{Sun:2020uoj} based on the AMPT model with a smooth transition from the partonic phase to the hadronic phase, which is around 0.4 regardless  of  the collision energy.  Our results thus suggest the observed non-monotonic behavior of \pdt, in particularly its enhancement seen in experiment measurements, could be due to the occurrence of a first-order phase transition between QGP and hadronic matter.

This paper is organized as follows. In Sec. II, we give a detailed description of the model setup for the present study,  which includes the 3-flavor NJL model and the partonic transport model derived from this model as well as the nucleon coalescence model used for producing deuteron and triton from the kinetically freeze-out nucleons. Results on the time evolution of the net-baryon density fluctuations and the yield ratio $N_{\rm t}N_{\rm p}/N_{\rm d}^2$ are then given in Sec. III for both cases with and without a first-order phase transition in the equation of state of the quark matter. We then present the conclusions and outlook in Sec. IV. Finally, an appendix is included to show that the effect of density fluctuations  on the yield ratio $N_{\rm t}N_{\rm p}/N_{\rm d}^2$ in the thermal model is similar to that in the coalescence model.

\section{Model Setup}

\subsection{Partonic interactions}

To describe the partonic  matter produced in relativistic heavy ion collisions, we follow Ref.~\cite{Li:2016uvu} by adopting the 3-flavor NJL model~\cite{Buballa:2003qv}  in terms of  the following Lagrangian density,
\begin{eqnarray}
 \mathcal{L}&=&\mathcal{L}_0 + \mathcal{L}_{\text{S}}+ \mathcal{L}_{\text{V}} + \mathcal{L}_{\text{det}},
\end{eqnarray}
with
\begin{eqnarray}
 \mathcal{L}_0 &=& \bar{\psi}(i\gamma^\mu \partial_\mu - \hat{m})\psi ,\notag  \\
 \mathcal{L}_{\text{S}} &=& G_{S}\sum_{a=0}^8 [(\bar{\psi}\lambda^a\psi)^2 + (\bar{\psi}i \gamma_5 \lambda^a \psi)^2], \notag\\
 \mathcal{L}_{\text{V}}&=& -g_V(\bar{\psi}\gamma^\mu\psi)^2 ,  \notag\\
 \mathcal{L}_{\text{det}} &=& -K[\text{det}\bar{\psi}(1+\gamma_5)\psi + \text{det}\bar{\psi}(1-\gamma_5)\psi],
\end{eqnarray}
where $\mathcal{L}_0$ is the non-interacting Lagrangian density for free quarks, $\psi = (u,d,s)^T$ represents the 3-flavor quark fields, and $\hat{m}=\text{diag}(m_u,m_d,m_s)$ is the current quark mass matrix. In the above equation, $\mathcal{L}_{\text{S}}$ is the interaction Lagrangian density for the quark scalar  interaction with $G_{S}$ being the scalar coupling constant,   $\lambda ^a $(a=1,...,8)  being the Gell-Mann matrices, and $\lambda^0 =\sqrt{2/3}I$. The term $\mathcal{L}_{\text{det}}$ is the Lagrangian density for the Kobayashi-Maskawa-t'Hooft (KMT) interaction~\cite{Hof78} that breaks the $U(1)_A$ symmetry with `det' denoting the determinant in flavor space~\cite{Bub05}, i.e.,
\begin{eqnarray}
 \text{det} (\bar{\psi} \Gamma \psi) = \sum_{i,j,k}\epsilon_\text{ijk}(\bar{u}\Gamma q_i)(\bar{d}\Gamma q_j)(\bar{s}\Gamma q_k).
\end{eqnarray}
This term gives rise to the six-point interactions in three  flavors. For the case of two flavors, the sum of scalar and psudo-scalar interactions and the KMT interaction with $K=-G_S$ reduces to the original NJL model. The term $\mathcal{L}_{\text{V}}$ represents the Lagrangian density for  the flavor-independent vector interaction with $g_{V}$ being the coupling strength~\cite{Mas13}.

As usually adopted in the application of the NJL model, we use the mean-field approximation~\cite{Wei91} to linearize the interactions, and the Lagrangian density becomes
\begin{eqnarray}\label{mf}
\mathcal{L} &=& \bar{u}(\gamma^\mu iD_{u\mu} - M_u)u+\bar{d}(\gamma^\mu iD_{d\mu} - M_d)d \notag \\
&&+ \bar{s}(\gamma^\mu iD_{s\mu} - M_s)s - 2 G_S(\phi_u^2+\phi_d^2+\phi_s^2) \notag \\
&&+ 4 K \phi_u\phi_d\phi_s+g_V(j_u^\mu+j_d^\mu+j_s^\mu)(j_{u\mu}+j_{d\mu}+j_{s\mu}),\notag \\
\end{eqnarray}
where
\begin{eqnarray}\label{eq:mass}
M_{u}&=& m_u - 4G_S\phi_u + 2K\phi_d\phi_s, \notag\\
M_{d}&=& m_d - 4G_S\phi_d + 2K\phi_u\phi_s, \notag\\
M_{s}&=& m_s - 4G_S\phi_s + 2K\phi_u\phi_d
\end{eqnarray}
are the in-medium effective masses of  $u$, $d$, and $s$ quarks, respectively, with $\phi_u=\langle\bar{u}u\rangle,~\phi_d=\langle\bar{d}d\rangle,~\text{and}~\phi_s=\langle\bar{s}s\rangle$ being their respective condensates.  In  Eq.~(\ref{mf}), $j_{u\mu}=\langle\bar{u}\gamma_\mu u\rangle,~j_{d\mu}=\langle\bar{d}\gamma_\mu d\rangle,~\text{and}~j_{s\mu}=\langle\bar{s}\gamma_\mu s\rangle $ denote, respectively, the u, d, and s net-quark vector currents, and
\begin{eqnarray}
iD_{u\mu}&=&i\partial_\mu - A_{u\mu},~~iD_{d\mu}=i\partial_\mu - A_{d\mu}, \notag\\
iD_{s\mu}&=&i\partial_\mu - A_{s\mu},
\end{eqnarray}
are the covariant derivatives, where the effective vector potentials are
\begin{eqnarray}\label{vector}
A_{u\mu}= A_{d\mu}=A_{s\mu}=2 g_V(j_{u\mu}+j_{d\mu}+j_{s\mu}).
\end{eqnarray}

\subsection{The equation of state}

The thermodynamic properties of a 3-flavor quark system are determined by the partition function $\mathcal{Z}$ with the path integral representation,
\begin{eqnarray}
\mathcal{Z} = \text{Tr} e^{-\frac{1}{T}(\hat{H}-\mu \hat{N})}=\int D\psi D\bar{\psi} e^{\int_0^\beta \int \mathcal{L} d\tau d^3x},
\end{eqnarray}
where $\beta=1/T$ and $\hat{H}$ are, respectively, the inverse of the temperature $T$ and the Hamiltonian operator, and $\mu$ and $\hat{N}$ are the chemical potential and  corresponding conserved charge number operator, respectively. The thermodynamic potential of   the quark matter of volume $V$ then reads
\begin{eqnarray}
\Omega &=& -\frac{T}{V} \text{ln} \mathcal{Z} = \Omega_u + \Omega_d+\Omega_s + 2 G_S(\phi_u^2+\phi_d^2+\phi_s^2)\notag \\
&&- 4 K \phi_u\phi_d\phi_s-g_V(\rho_{u}+\rho_{d}+\rho_{s} )^2,
\end{eqnarray}
where $\rho_{f}$ is the net-quark number density of quarks of flavor $f=u,s,d$, and
\begin{eqnarray}
\Omega_f &=& -2N_c \int \frac{d^3p}{(2\pi)^3} \left[E_f + T \text{ln}\left(1+e^{\frac{E_f - \mu^*_f}{T}}\right) \right.\notag \\
         &&\left.+T \text{ln}\left(1+e^{\frac{E_f + \mu^*_f}{T}}\right)\right],
\end{eqnarray}
with $E_f=(M_f^2+{ p}^2)^{1/2}$ being  the  quark  energy and
\begin{eqnarray}
\mu_u^*&=&\mu_u-2 g_V(\rho_{u}+\rho_{d}+\rho_{s}), \notag \\
\mu_d^*&=&\mu_d-2 g_V(\rho_{u}+\rho_{d}+\rho_{s}), \notag \\
\mu_s^*&=&\mu_s-2 g_V(\rho_{u}+\rho_{d}+\rho_{s}),
\end{eqnarray}
being the effective chemical potentials.

Minimizing the thermodynamical potential with respect to the quark in-medium masses and chemical potentials, i.e.,
\begin{eqnarray}
\frac{\delta\Omega}{\delta M_f}  = \frac{\delta\Omega}{\delta \mu^*_f}  =0 ,
\end{eqnarray}
leads to the  following  expressions for the quark condensates and the net-quark number densities,
\begin{eqnarray}
\phi_f &=& 2N_c \int \frac{d^3p}{(2\pi)^3} \frac{M_f}{E_f} (n_{f+}+ n_{f-}-1), \label{Eq13} \notag\\
\rho_f &=& 2N_c \int \frac{d^3p}{(2\pi)^3}(n_{f+}-n_{f-}),\label{Eq14}
\end{eqnarray}
where the occupation numbers are $n_{f\pm} = \left(e^{\frac{E_f\mp\mu_f^*}{T}}+ 1\right)^{-1}$ with the upper (lower) sign for quarks (antiquarks). The net-quark number density $\rho_f$ is the time component of the net-quark   vector  current
\begin{equation}
j_{f\mu} = 2N_c \int \frac{d^3p}{(2\pi)^3} \frac{p_\mu}{E_f}(n_{f+}-n_{f-}),
\end{equation}
whose space components vanish in thermodynamical calculations for a static system, but become effective in transport simulations.  Because the NJL model is unrenormalizable, a momentum cutoff $\Lambda$ is usually introduced in the integrals in above equations.

\begin{figure}[h]
  \centering
  \includegraphics[width=0.5\textwidth]{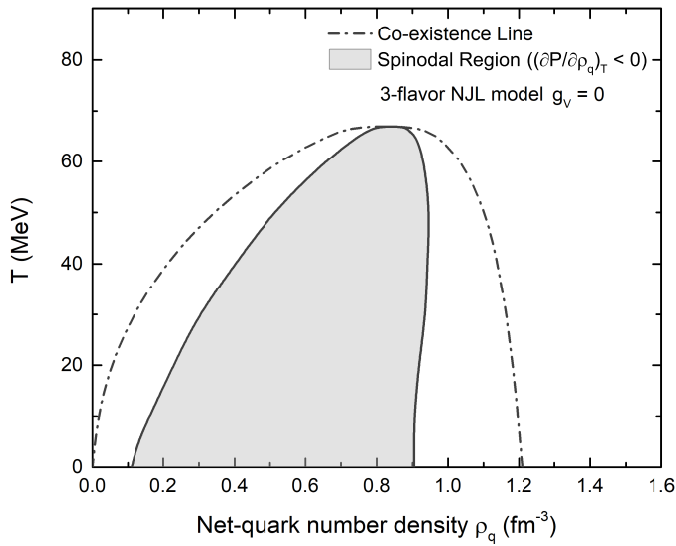}
  \caption{\protect  Quark matter phase diagram in the temperature and the net-quark density plane from the three-flavor NJL model with $g_V=0$ with equal chemical potential for $u$ and $d$ quarks and zero chemical potential for $s$ quarks.}
  \label{pic1}
\end{figure}

For the parameters in the NJL model, we use the values $m_u=m_d=5.5$~MeV, $m_s=140.7$~MeV, $G_S\Lambda^2=1.835$, $K\Lambda^5=12.36$, and  $\Lambda=602.3$~MeV~\cite{Bra12,Lut92,Buballa:2003qv}.  Figure~\ref{pic1} shows the quark matter phase digram in the temperature and density plane  for the case of vanishing vector coupling $g_V=0$.  The dash-dotted line denotes the coexistence line for the chiral symmetry restored and broken phases, while the  solid line denotes the line on which the condition $(\partial P/\partial \rho_q)_T=0$ is satisfied, where $\rho_q = \rho_u+\rho_d+\rho_s$ is the net-quark number density and the pressure is  given by $P=-\Omega+\Omega_0$ with $\Omega_0$ being  the thermodynamical potential  of the QCD vacuum. For quark matter in between these two lines, it is in the metastable phase.    In the shaded region,  where the quark matter  has a convex anomaly in  its  pressure $(\partial P/\partial \rho_q)_T<0$  and is thus  unstable against phase decompositions~\cite{Chomaz:2003dz}. The critical temperature and net-quark density in this case are around 70 MeV and 0.85 fm$^{-3}$, respectively. We note that the critical temperature in the NJL model is smaller than those obtained from the  approach based on the Schwinger-Dyson equation~\cite{Xin:2014ela,Fischer:2014ata} and the functional renormalization group method~\cite{Fu:2019hdw,Gao:2020qsj}. Including the repulsive vector interaction of coupling $g_V=G_S$, the critical point disappears in the phase diagram,  and  the transition from the chirally restored phase to the broken phase changes to a smooth crossover.

\subsection{Equations of motion in transport simulations}

To construct a partonic transport model from the NJL model, one needs the equations of motion of quarks and antiquarks (partons)  of a given flavor  in their mean fields. They can be obtained from the effective vector mean-field potential $A^\mu = (A_0,{\bf A})$  in  Eq.~(\ref{vector}) and the in-medium quark mass   in Eq.~(\ref{eq:mass})  through the single parton Hamiltonian,
\begin{eqnarray}\label{hamiltonian}
H_{\pm} = \pm A_0 + \sqrt{M^2 + ( {\bf P \mp A})^2},
\end{eqnarray}
where the upper (lower) sign  is  for quarks (antiquarks), and ${\bf P}={\bf p}\pm{\bf A}$  is the canonical momentum with ${\bf p}$ being the kinetic momentum.  

In terms of the strong electric and magnetic fields, given by $\bf E=-\frac{\partial{\bf A}}{\partial t}+\boldsymbol\nabla_r {{\rm A}_0}$ and $\bf B=\boldsymbol\nabla_r\times{\bf A}$, respectively, and the effective energy of a quark $E^* = \sqrt{M^2+{\bf p}^2}$, the time evolution of the partonic matter produced in relativistic heavy-ion collisions can be described by  a transport equation for the parton phase-space distribution functions $f_{\pm}(\vec{r},\vec{p},t)$, similar to  that for the nucleonic  matter based on the Walecka model~\cite{Ko87,Ko88}, i.e.,
\begin{eqnarray}\label{transport}
&&\frac{\partial f_{\pm}}{\partial t} + {\bf v}\cdot \boldsymbol\nabla_r f_{\pm} +  \left(-\frac{M}{E^*} \boldsymbol\nabla_r M \pm  {\bf E} \pm {\bf v}\times {\bf B}\right)\cdot \nabla_p f_{\pm}  \notag\\
&&\qquad= \left(\frac{\partial f_{\pm}}{\partial t}\right)_{\rm coll}
\end{eqnarray}
with ${\bf v}=\frac{d{\bf r}}{dt}=\frac{{\bf p}}{E^*}$. In the above,   $(\frac{\partial f_{\pm}}{\partial t})_{\rm coll}$ is the collision term for describing the change of the phase-space distribution function of partons due to their scatterings and is given by the integral
\begin{eqnarray}
&&\left(\frac{\partial f_{\pm}}{\partial t}\right)_{\rm coll}=-\frac{1}{(2\pi)^3}\int\int d{\bf p}_2d{\bf p}_3\int d\Omega|v_{12}|\frac{d\sigma}{d\Omega}\notag\\
&&\qquad\times\delta^{(3)}({\bf p}+{\bf p}_2-{\bf p}_3-{\bf p}_4)\{f_\pm({\bf r},{\bf p},t)f({\bf r},{\bf p}_2,t)\notag\\
&&\qquad\times[2N_c-f({\bf r},{\bf p}_3,t)][2N_c-f({\bf r},{\bf p}_4,t)]\notag\\
&&\qquad-f({\bf r},{\bf p}_3,t)f({\bf r},{\bf p}_4,t)\notag\\
&&\qquad\times[2N_c-f_\pm({\bf r},{\bf p},t)][2N_c-f({\bf r},{\bf p}_2,t)]\},
\end{eqnarray}
where $v_{12}$ is the relative velocity between the two scattering partons and $d\sigma/d\Omega$ is  their differential scattering cross section. The equilibrated parton distributions from the transport model   are   $f_\pm = 2N_c n_\pm$, where $n_\pm$   are  the occupation  probabilities defined below   Eq.~(\ref{Eq13})  with the flavor index  omitted.

Solving the transport equation by the test particle method~\cite{Won82} leads to the following equations  of motion for a test parton between its scattering with another parton:
\begin{eqnarray}
\frac{d\bf r}{d\bf t}  &=&\bf v,\notag\\
\frac{d{\bf p}}{dt}&=& -\frac{M}{E^*} \boldsymbol\nabla_r M\pm{\bf E}\pm \bf{v}\times {\bf B}.
\end{eqnarray}
In the test particle method, the parton scalar and vector densities are determined by dividing the  system into cells  in space and including in each cell only test partons that have momenta in the rest frame of the cell  less than the cutoff  momentum $\Lambda$. The in-medium quark masses and the scalar field are calculated in the cell rest frame, and the vector fields are calculated in the laboratory frame with corresponding coupling constants. For test partons with momentum above the cutoff, they are not affected by the mean fields and are thus treated as free particles.  For the collision term in the transport equation in Eq.~(\ref{transport}), only partons in the same physical event are allowed to scatter with each other and  the geometrical method of Ref.~\cite{Bertsch:1988ik} is used to treat their scatterings.

\subsection{Density fluctuations and light nuclei production}

To quantify the density fluctuation, we consider the scaled density moments $y_N=\overline{\rho^N}/\overline{ \rho}^N$~\cite{Steinheimer:2012gc} with
\begin{eqnarray}\label{Eq3-1}
\overline{\rho^N} =  \frac{\int d{\bf x}\rho^{(N+1)}({\bf x})}{\int d{\bf x}\rho({\bf x})}.
\end{eqnarray}
It is easy to show that the first-order scaled density moment $y_1 = 1$ and that $y_N=1$ if the density is a constant.  In the present study, we are interested in the second-order scaled density moment, i.e.,
\begin{eqnarray}
y_2 = \frac{[\int d{\bf x}\rho({\bf x})][\int \text{d}\bf x\rho^3(\bf{x})]}{[\int \text{d}\bf x\rho^2(\bf x)]^{2}}. \label{Eq3-2}
\end{eqnarray}
For small density fluctuations, $\rho(\bf x) = \rho_0+\delta \rho(\bf x)$ with $\rho_0$ being the   average density, one can obtain the following relation
\begin{eqnarray}
y_2 \approx 1+\frac{\int \text{d}\bf x(\delta\rho(\bf x))^{2}}{\int \text{d}\bf x\rho_0^{2}}\equiv 1+\Delta \rho. \label{Eq3-3}
\end{eqnarray}
where $\Delta\rho$ is defined as the relative density fluctuation averaged over space~\cite{Sun:2017xrx,Sun:2018jhg}.

Although the density fluctuation in heavy ion collisions is not directly accessible in experimental measurements, it can affect the production of light nuclei as  pointed out in Refs.~\cite{Sun:2017xrx,Sun:2018jhg}.   Recently,   understanding light nuclei production in relativistic heavy ion collisions has attracted a lot of theoretical interest~\cite{Ma:2013xn,Andronic:2017pug,Manju:2018knm,Braun-Munzinger:2018hat,Chen:2018tnh,Bazak:2018hgl,Zhao:2018lyf,Dong:2018cye,Bellini:2018epz,Sun:2018mqq,Oliinychenko:2018ugs,Xu:2018jff,Cai:2019jtk,Mrowczynski:2019yrr,Kachelriess:2019taq,Blum:2019suo,Oliinychenko:2020ply,Mrowczynski:2020ugu,Donigus:2020ohq,Shao:2020lbq,Vovchenko:2020dmv,Ye:2020lrc,Bazak:2020wjn,Liu:2019nii,Sun:2020uoj}. To illustrate the relation between density fluctuations and light nuclei production, we adopt the nucleon coalescence model to study light nuclei production  using kinetic freeze-out nucleons  in heavy ion collisions from a simple fireball model in this section and then  from a more realistic transport model in the next section.

For deuteron production from an emission source of protons and neutrons, its number in  the coalescence model is calculated from the overlap of the proton and neutron phase-space distribution functions $f_{p,n}({\bf x},{\bf p})$ with the Wigner function $W_{\rm d}({\bf x},{\bf p})$ of the deuteron internal wave function~\cite{Gyulassy:1982pe}, i.e.,
\begin{eqnarray}
N_\text{d}&=&g_\text{d}\int \text{d}^3{\bf x}_1  \text{d}^3{\bf p}_1  \text{d}^3{\bf x}_2 \text{d}^3{\bf p}_2 f_n({\bf x}_1,{\bf p}_1) \notag \\
&&\times f_p({\bf x}_2,{\bf p}_2)W_\text{d}({\bf x},{\bf p}), \label{Eq3-4}
\end{eqnarray}
with $g_d=3/4$ being the statistical factor for spin 1/2 proton and neutron to from a spin 1 deuteron.  For the Wigner function of the deuteron, which is given by the Wigner transform of its internal wave function,   it is taken to be
\begin{eqnarray}
W_\text{d}({\bf x},{\bf p})=8\exp\left({-\frac{x^2}{\sigma^2}}-\sigma^2 p^2\right), \label{Eq3-5}
\end{eqnarray}
by  using  a Gaussian or harmonic oscillator wave function for its internal wave function as usually assumed in the coalescence model for deuteron production. Normalizing the deuteron Wigner function  according to $\int \text{d}^3{\bf x}\int \text{d}^3{\bf p}~W_d({\bf x},{\bf p})=(2\pi)^3$, it gives the probability for a proton and a neutron that are separated by the relative coordinate  ${\bf x}$ and relative momentum ${\bf p}$  to form a deuteron.  Together with the center-of-mass coordinate ${\bf X}$ and momentum ${\bf P}$ of the coalescing proton and neutron, they are defined by
 \begin{eqnarray}
{\bf X}=\frac{{\bf x}_1+{\bf x}_2}{2},\quad {\bf x}=\frac{{\bf x}_1-{\bf x}_2}{\sqrt{2}}, \notag \\
{\bf P}={\bf p}_1+{\bf p}_2,\quad{\bf p}=\frac{{\bf p}_1-{\bf p}_2}{\sqrt{2}}. \label{Eq.Jacobi}
\end{eqnarray}
The parameter $\sigma$ in the deuteron Wigner function (Eq. (\ref{Eq3-5})) is related  to  the root-mean-square radius $r_{\text d}$ of deuteron through $\sigma  = \sqrt{4/3}~r_{\text d}\approx 2.26$ fm~\cite{Ropke:2008qk,Sun:2015jta,Sun:2017ooe} and is much smaller than the size of the nucleon emission source  considered in the present study.  We note that the    coordinate transformations  in Eq.~(\ref{Eq.Jacobi})   conserve the volume in phase space, instead of  the volumes in coordinate and momentum spaces separately.

For protons and neutrons emitted from a locally thermalized fireball of   temperature $T$ and volume $V$, their distribution functions in the non-relativistic approximation are   given by $f_{p,n}({\bf x},{\bf p})=\rho_{p,n}({\bf x})(2\pi mT)^{-\frac{3}{2}}~e^{-\frac{p^2}{2mT}}$, where $m$ is the nucleon mass and $\rho_{p,n}(\bf x)$ is the local proton or neutron density.  In this case, the  integral in Eq.~(\ref{Eq3-4}) can be straightforwardly evaluated, leading to
\begin{eqnarray}
N_\text{d}&=&\frac{2^{3/2} g_\text{d}}{(2\pi mT)^3}\int \text{d}^3{\bf x_1}  \text{d}^3{\bf x_2}~\rho_\text{n}(\bf{x_1})\rho_\text{p}(\bf{x_2})e^{-\frac{(\bf{x_1}-\bf{x_2})^2}{2\sigma^2}}  \notag \\
 &&{\times}\int \text{d}^3{\bf P}~e^{-\frac{P^2}{4mT}}  \int \text{d}^3{\bf p}~e^{-p^2(\sigma^2+\frac{1}{2mT})}\nonumber\\
&\approx&\frac{3}{2^{1/2}}\left(\frac{2\pi}{mT}\right)^{3/2}\int \text{d}^3{\bf x_1}   \text{d}^3{\bf x_2}\rho_n({\bf x_1})\rho_p({\bf x_2})  \nonumber\\
&&{\times}\frac{1}{(2\pi \sigma^2)^{3/2}}e^{-\frac{({\bf x_1-x_2})^2}{2\sigma^2}}, \label{Eq3-6}
\end{eqnarray}
where the second expression follows from the fact that the thermal wavelength of a nucleon  in the kinetically freeze-out  hadonic matter is much smaller than the size of a deuteron.

Similarly, the number of tritons from the coalescence of two neutrons and one proton is given by
\begin{eqnarray}
N_\text{t}&=&g_\text{t}\int \text{d}^3{\bf x}_1  \text{d}^3{\bf p}_1  \text{d}^3{\bf x}_2 \text{d}^3{\bf p}_2\text{d}^3{\bf x}_3 \text{d}^3{\bf p}_3 f_n({\bf x}_1,{\bf p}_1) \notag \\
&&f_n({\bf x}_2,{\bf p}_2)f_p({\bf x}_3,{\bf p}_3)W_\text{t}({\bf x},\boldsymbol{\lambda},{\bf{p}},{\bf{p}_\lambda}), \label{Eq3-6-1}
\end{eqnarray}
where $g_\text{t}=1/4$ is the statistical factor for two spin 1/2 neutrons and one spin 1/2 proton to form a spin 1/2 triton.  The   triton Wigner function is  taken to be
\begin{eqnarray}
W_\text{t}({\bf x},{\boldsymbol{ \lambda}},{\bf p},{\bf p_\lambda})=8^2\exp\left({-\frac{x^2}{\sigma_\text{t}^2}-\frac{\lambda^2}{\sigma_\text{t}^2}}-\sigma_\text{t}^2 p^2-\sigma_\text{t}^2 p_\lambda^2\right),\notag \\\label{Eq3-6-2}
\end{eqnarray}
with the additional relative coordinate $\boldsymbol\lambda$ and relative momentum ${\bf p}_\lambda$ defined together with the center-of-mass coordinate ${\bf X}$ and momentum ${\bf P}$ of the three coalescing nucleons by~\cite{Chen:2003ava,Sun:2015jta,Sun:2017ooe}
 \begin{eqnarray}
{\bf X}=\frac{{\bf x}_1+{\bf x}_2+{\bf x}_3}{3},\quad{\boldsymbol\lambda}=\frac{{\bf x}_1+{\bf x}_2-2{\bf x}_3}{\sqrt{6}}, \notag \\
{\bf P}={\bf p}_1+{\bf p}_2+{\bf p}_3,\quad{\bf p_\lambda}=\frac{{\bf p}_1+{\bf p}_2-2{\bf p}_3}{\sqrt{6}}.\notag\\
\end{eqnarray}
The parameter $\sigma_\text{t}$ in the triton Wigner function (Eq.~(\ref{Eq3-6-2})) is related to the root-mean-square radius $r_\text{t}$ of triton through $\sigma_\text{t}=r_\text{t}=1.59$ fm~\cite{Sun:2015jta,Sun:2017ooe,Ropke:2008qk}. The  number of  tritons produced from the above thermal emission source is then
\begin{eqnarray}
N_\text{t}&\approx&\frac{3^{3/2}}{4}\left(\frac{2\pi}{mT}\right)^{3}\int \text{d}^3{\bf{x_1}}   \text{d}^3{\bf{x_2}} \text{d}^3 {\bf{x_3}}\rho_n({\bf{x_1}})\rho_n({\bf{x_2}}) \nonumber\\
&&{\times}\rho_p({\bf{x_3}}) \frac{1}{3^{3/2}(\pi \sigma_\text{t}^2)^{3}}e^{-\frac{({\bf{x_1}}-{\bf{x_2}})^2}{2\sigma_\text{t}^2}-\frac{({\bf{x_1}}+{\bf{x_2}}-2{\bf{x_3}})^2}{6\sigma_\text{t}^2}},  \label{Eq3-7}
\end{eqnarray}
after taking into consideration of the much smaller nucleon thermal wavelength than the triton size.

For a nucleon distribution with density fluctuations over  a length scale much larger than the sizes of deuteron and triton, one can neglect higher-order correlations in density fluctuations and rewrite Eqs.~(\ref{Eq3-6}) and (\ref{Eq3-7})  as~\cite{Sun:2017xrx,Sun:2018jhg}
\begin{eqnarray}
N_{\rm d} &\approx& \frac{3}{2^{1/2}}\left(\frac{2\pi}{mT}\right)^{3/2} \int \text{d}^3{\bf x} ~\rho_n({\bf x})\rho_p({\bf x}) \notag \\
&\approx& \frac{3}{2^{1/2}}\left(\frac{2\pi}{mT}\right)^{3/2} N_p\langle\rho_n\rangle(1+C_\text{np}) . \label{Eq3-8}
\end{eqnarray}
and
\begin{eqnarray}
N_\text{t} &\approx& \frac{3^{3/2}}{4}\left(\frac{2\pi}{mT}\right)^{3} \int \text{d}^3{\bf x} ~\rho_n^2({\bf x})\rho_p({\bf x})\notag \\
&\approx& \frac{3^{3/2}}{4}\left(\frac{2\pi}{mT}\right)^{3} N_p\langle\rho_n\rangle^2(1+\Delta\rho_n+2C_\text{np}).  \label{Eq3-9}
\end{eqnarray}
In the above, $C_\text{np}=\langle\delta\rho_n(x)\delta\rho_p(x)\rangle/ (\langle\rho_n\rangle\langle\rho_p\rangle)$ denotes the correlation between the neutron and proton density fluctuations and $\Delta\rho_n=\langle\delta\rho_n(x)^2\rangle/(\langle\rho_n\rangle^2)$ is the relative neutron density fluctuation as defined in Eq.~(\ref{Eq3-3}), where $\langle\cdots\rangle$ denotes the average over  the coordinate  space ~\cite{Sun:2017xrx,Sun:2018jhg}.

The yield ratio \pdt is then given by
\begin{eqnarray}
\mathcal{O}_{\text{p-d-t}}&=& \frac{N_\text{t}N_\text{p}}{N_\text{d}^2} \approx\frac{1}{2\sqrt{3}}  \frac{1+2C_\text{np}+\Delta\rho_n}{(1+C_\text{np})^2}.  \label{Eq3-10}
\end{eqnarray}
In the case that the correlation between the neutron and proton density fluctuations $C_\text{np}$ is small, one has in the leading-order approximation,
\begin{eqnarray}
\mathcal{O}_{\text{p-d-t}}&=& \frac{N_\text{t}N_\text{p}}{N_\text{d}^2}\approx\frac{1}{2\sqrt{3}}(1+\Delta\rho_n) \approx\frac{1}{2\sqrt{3}}y_2, \label{Eq3-11}
\end{eqnarray}
where the last step follows from the definition  of the second-order  scaled neutron density moment   in Eq.~(\ref{Eq3-2}).  The above equation shows that the information on spatial density fluctuations is  encoded in the  yield ratio   $N_tN_p/N_d^2$, and a large neutron density fluctuation would lead to an enhancement of its value. This is different from the net-baryon density fluctuation, which is  related to the susceptibility of conserved baryon charge~\cite{Luo:2017faz}. However, since nucleons carry most of the baryon charges in heavy-ion collisions at energies with significant stoppings, the nucleon density fluctuation   is expected to be  closely related to the (net-)baryon density fluctuation.    The above effect of  nucleon density fluctuations  on deuteron and triton production  is also expected to affect the production of heavier nuclei   such as the  $^4$He~\cite{Sun:2017xrx,Shuryak:2019ikv,Shuryak:2020}.  The  relation between the nucleon density fluctuation and the yield ratio \pdt given by Eq.~(\ref{Eq3-11}) is quite general and is not restricted to the coalescence model considered here.  As shown in the Appendix, they can also be obtained in the thermal model  that assumes local thermal and chemical equilibrium  among nucleons, deuterons and tritons with a spatially varied  chemical potential for nucleons.

\section{\protect Results}

\begin{figure*}[t]
  \centering
  \includegraphics[width=1.5\textwidth, bb=60 20 1200 300]{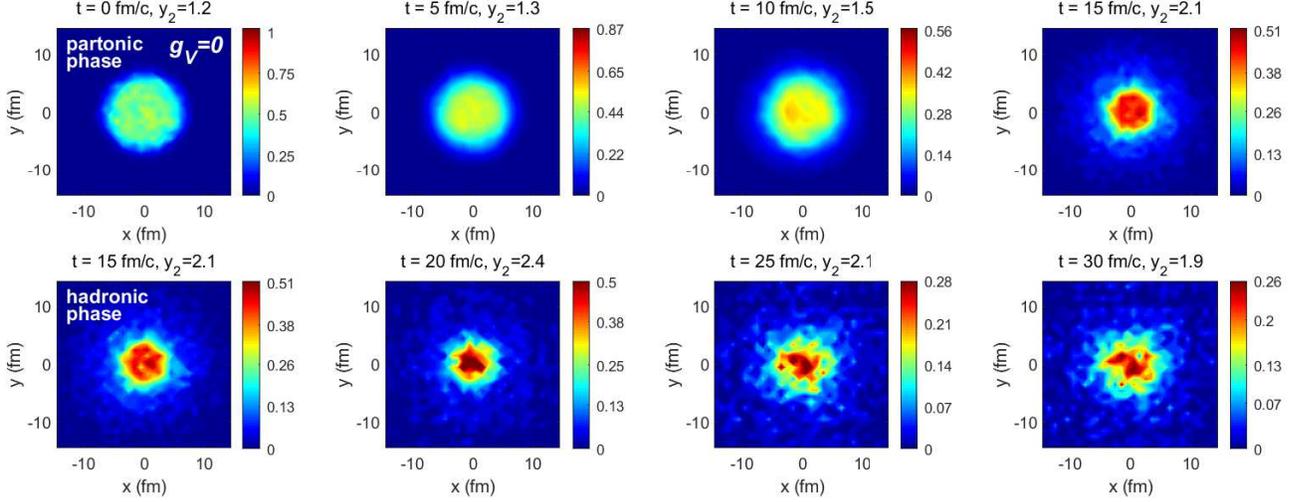}
  \caption{\protect Evolution of the  distribution of net-baryon number density in fm$^{-3}$  and the second-order scaled density moment $y_2$ (shown above individual  windows) in transverse plane $z=0$ in the partonic phase (upper  windows) and the hadronic phase (lower windows) for the case of $g_V = 0$, corresponding to a first-order quark to hadronic matter transition. }
  \label{pic2}
\end{figure*}

To illustrate the effect of a first-order quark to hadronic matter phase transition on light nuclei production in relativistic heavy ion collisions, we take the initial distribution of quarks and antiquarks in  the coordinate  space  as in Ref.~\cite{Li:2016uvu} by letting them  to follow a spherical Woods-Saxon form:
\begin{eqnarray}
\rho(r) = \frac{\rho_0}{1+\exp((r-R)/a)},\label{Eq.Wood}
\end{eqnarray}
with a radius $R=6$~fm, a surface thickness parameter $a = 0.6$ fm, and a central net-quark density of $\rho_0=$~1.5 fm$^{-3}$. The momenta of these partons are taken to follow a relativistic Boltzmann distribution with the temperature $T=70$ MeV and the chemical potentials  $\mu_u=\mu_d$ and $\mu_s=0$.  For the equation of state of the partonic matter, we consider the two cases with and without a first-order chiral phase transition and a critical endpoint by setting the vector coupling constant in the NJL model to $g_V=0$  and $g_V=G_S$, respectively. In both cases, an isotropic cross section of 3 mb is used for the scattering between two partons, and the Pauli-blocking factors in the collision term of the transport equation are neglected as their effects are appreciable only in the very beginning of the time evolution of the system when the parton density is high and become unimportant as the system expands.

\subsection{\protect First-order chiral phase transition}

\begin{figure}[h]
  \centering
  \includegraphics[width=0.48\textwidth]{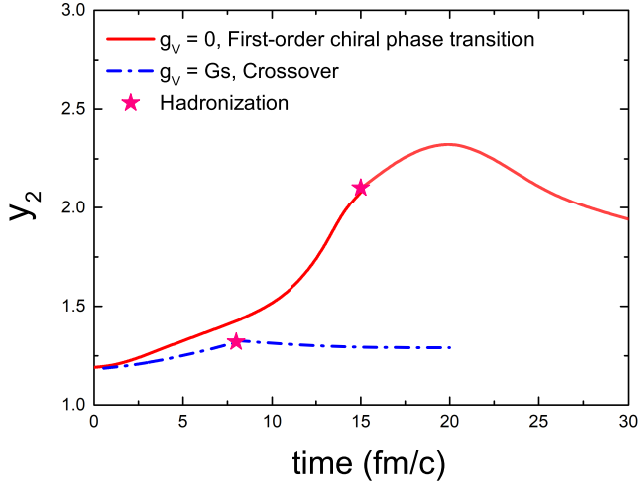}
  \caption{Time dependence of the second-order scaled  moment $y_2$ of net-baryon density for the two cases of $g_V=0$ with a first-order chiral phase transition (solid line) and $g_V=G_S$ with a smooth crossover  chiral transition (dash-dotted line).  Red solid stars indicate the time at which hadronization takes place. }
  \label{pic:y2}
\end{figure}

\begin{figure}[h]
  \centering
  \includegraphics[width=0.5\textwidth]{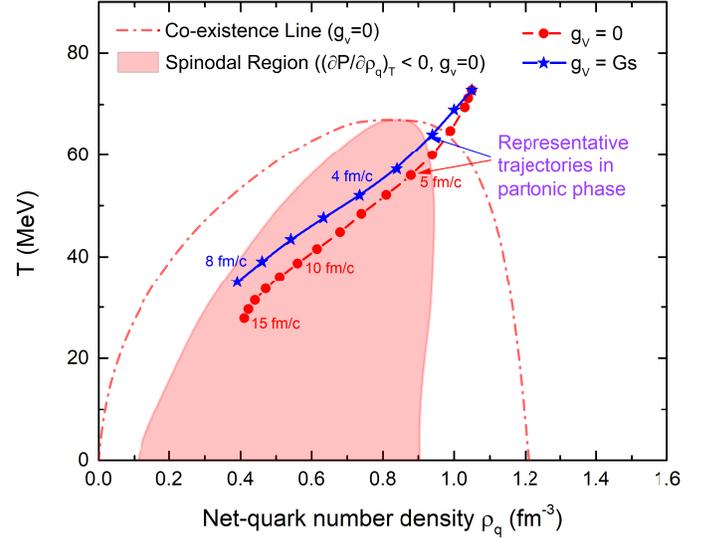}
  \caption{\protect Evolution trajectories of the partonic matter in the plane of temperature $T$ versus net quark number density $\rho_q$ for the two cases of $g_V=0$ with a first-order  chiral  phase transition (red solid circles) and $g_V=G_S$ with a crossover  chiral transition  (solid stars). The spinodal region in the case of $g_V=0$ is denoted by the shaded region. }
  \label{pic:trajectory}
\end{figure}

We first  consider the case of vanishing vector coupling constant $g_V = 0$, corresponding to a    partonic  matter that undergoes a first-order chiral phase transition. Shown in Fig. \ref{pic2} is the  distribution of net-baryon number density in fm$^{-3}$ in the transverse plane ($z=0$) during the time evolution of both the partonic phase (upper  windows) and the hadronic phase (lower  windows).   It is seen that the second-order scaled density moment $y_2$, given above each window  and also shown in Fig.~\ref{pic:y2} by the solid line, during the partonic evolution increases as the system expands. The{se} results   are obtained from 100 events with  100 test particles  used  in  each event to solve   the transport equation.

To understand the origin of the density fluctuations, we consider the evolution trajectory of the system shown by red solid circles in Fig.~\ref{pic:trajectory}. Each point on the trajectory is obtained from the average temperature and net-quark number density of the system over all its cells weighted by the parton number in each cell, with the temperature and net-quark number density in each cell determined from assuming the   energy density and the net-quark number density  in the rest frame of the cell being the same as those of a thermally equilibrated system. For the present case of $g_V=0$, the system enters the spinodal region of the phase diagram, shown by the shaded region, at around $t=$~4 fm/c. With decreasing temperature and density as the system expands, density fluctuations gradually develop as a result of the spinodal instability and reach a value of $y_2\sim 2$ at $t=15$~fm/$c$.    This value is larger than the value of 
$\sim 1.7$ in Ref.~\cite{Li:2016uvu}, and this is because a smearing density distribution $\rho(x)$ over neighboring cells is used in the present study to calculate the $y_2$ according Eq.(\ref{Eq3-3}).  We note that for    the initial conditions used in the present study, the scaled density moment  $y_2$ of net-baryon number density    has  a similar value as that of   the baryon  number density.

\begin{figure*}[t]
  \centering
  \includegraphics[width=1.5\textwidth, bb=60 20 1200 280]{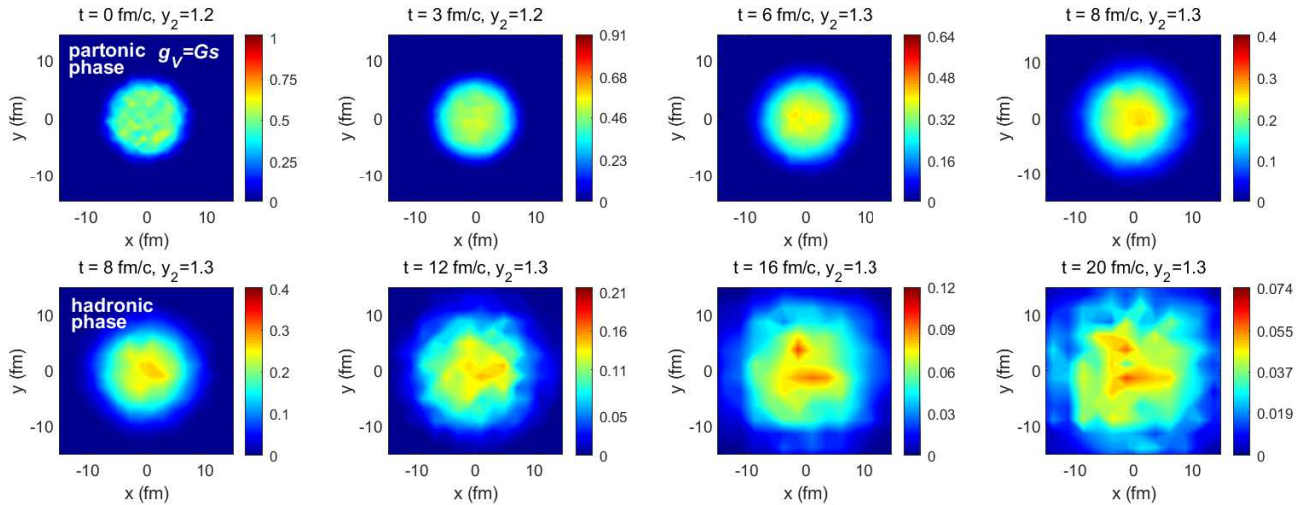}
\caption{\protect Same as Fig.~\ref{pic2}  for the case of $g_V = G_S$, corresponding to a smooth crossover from the quark to the hadronic matter.}
  \label{pic3}
\end{figure*}

Although the chiral symmetry broken matter in the coexistent region inside the dash-dotted line in Fig.~\ref{pic:trajectory} should in principle be described in terms of hadronic degrees of freedom if one assumes that the chiral phase transition and the deconfinement transition coincide, we use the NJL model to describe its properties and dynamics in the present study for simplicity.  In this respect, we adopt the condition for the hadronization of the    partonic  matter by choosing  the in-medium quark mass to have an average value of around 260 MeV, which is about 70 percent of its mass $M_{0}=367.7$ MeV in vacuum. This condition is fulfilled at $t_h=15$~fm/$c$ for the case of $g_V=0$, when we  convert all partons in the NJL transport model to hadrons through the spatial quark coalescence model in the AMPT~\cite{Lin05}.  The scatterings and propagations of these hadrons are then described by  the ART model~\cite{Li:1995pra} in AMPT.   Results on the time evolution of the net-baryon density distribution in the transverse plane at $z=0$ of the hadronic matter are shown in the lower panels of Fig.~\ref{pic2}.  It is seen that except a very small discontinuity in the value of $y_2$ at hadronization due to the spatial coalescence model used in AMPT to convert partons to hadrons, the net-baryon density distribution at hadronization is entirely carried over from the partonic phase to the hadronic phase, suggesting the conservation of local  baryon charge in the coalescence model used in the AMPT for hadronization. Also, the large second-order scaled net-baryon density moment $y_2$  at hadronization remains substantial during the hadronic evolution as shown in Fig.~\ref{pic:y2} by the solid line starting from the red star when hadronization occurs.

The above seemingly counter-intuitive result that the large density fluctuations developed at the phase transition can survive strong hadronic scatterings  is  due to the rapid expansion of the hadronic matter compared to the relative slow global thermal  equilibration of conserved charges through diffusion~\cite{Asakawa:2015ybt}.  This phenomenon is analogous to the  temperature fluctuations in  the cosmic microwave background (CMB) at different angles, which is now considered as the remnant of quantum fluctuations in the primordial Universe during the rapid inflation  epoch~\cite{Ade:2013sjv,Baumann:2009ds}. Also, according to Eq.~(\ref{Eq3-2}), the second-order scaled density moment $y_2$ would remain a constant if the expansion of the hadronic matter is self-similar, i.e., $\rho(\lambda(t) x,t) =  \alpha(t)\rho(x,t_h)$, where $\lambda$ and $\alpha$ are any real functions of time $t$, similar to the expansion of the Universe.

\subsection{ Smooth crossover}

For the case of $g_V=G_S$ with a smooth crossover chiral transition, results on the net-baryon number density distributions in the transverse plane during the evolution of  the partonic and hadronic phases are shown in the upper and lower windows  of Fig. \ref{pic3}, respectively.   The corresponding $y_2$ shown in Fig.~\ref{pic:y2} by the dash-dotted line is much smaller than the case with a first-order chiral phase transition and only changes slightly throughout the partonic and hadronic  evolutions of the system, as the density fluctuations in this case are largely due to  statistical fluctuations and the finite size of the system.  Also, the system in this case expands much faster than in the case of $g_V=0$ with a first-order phase transition as shown by the solid stars in Fig.~\ref{pic:trajectory}. A  similar consideration based on the quark in-medium mass as in the case with a first-order chiral phase transition leads to a global hadronization at $t_h=8$ fm/c.

We note that the difference in the behavior of the scaled density moment $y_2$ in Fig.~\ref{pic:y2} between the two cases is mainly due to whether there is a spinodal instability in the equation of state, not because of their different expansion rates.  This argument is consistent with the results in Refs.
~\cite{Steinheimer:2012gc,Steinheimer:2019iso}. Although  two equations of state with and without spinodal instabilities are also used in this study, they are related by the Maxwell construction and thus describe the same first-order phase transition.  It is found that these two equations of state give very different behaviors in the density moments as in our study,    but  lead to similar radial expansions for the produced matter in heavy ion collisions.

\subsection{Light nuclei production}

\begin{figure}[b]
  \centering
  \includegraphics[width=0.46\textwidth]{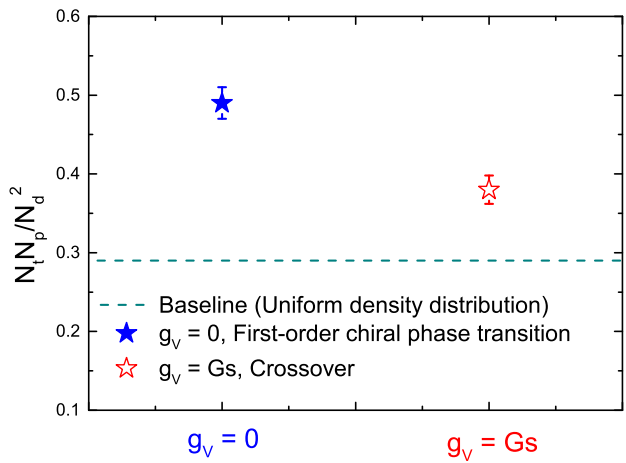}
  \caption{Yield ratio \pdt  of proton ($p$), deuteron ($d$), and triton ($t$) for the cases  of $g_V=0$ with a first-order phase transition (solid star) and $g_V=G_S$ with a crossover  transition (open star).  The dashed line denotes the value of about 0.29 for this ratio  when using a uniform density distribution  in Eq.~(\ref{Eq3-11}).}
  \label{pic5}
\end{figure}

Although spatial density fluctuations cannot be directly measured in experiments, it can lead to an enhancement~\cite{Sun:2017xrx,Sun:2018jhg} of the yield ratio $ N_\text{t}N_\text{p}/ N_\text{d}^2$  as seen from Eq.~(\ref{Eq3-11}). Using the phase-space distribution of nucleons at the kinetic freeze out in AMPT, we can calculate the production probabilities of deuteron and triton according to Eq. (\ref{Eq3-5}) and Eq. (\ref{Eq3-6-2}), respectively.   To achieve good statistics, test  nucleons   in each event are allowed to coalesce to deuterons and tritons.  As shown in Fig.~\ref{pic5},  the yield ratio $ N_\text{t}N_\text{p}/N_\text{d}^2$ is  about $0.49\pm0.02$ (solid star) for the case of a first-order chiral phase transition and is about $0.38\pm 0.02$  (open star) for the case of a smooth crossover transition.  Compared with the value of about 0.29, given by  Eq.~(\ref{Eq3-11})  for a uniform density distribution, shown by the dashed line in Fig.~\ref{pic5}, the enhancements of the yield ratio $ N_\text{t}N_\text{p}/ N_\text{d}^2$ in these two cases are about 1.7 and 1.3, respectively.   These values of enhancement are similar to those expected from the values of $y_2$ in Fig.~\ref{pic:y2}, which, according to Eq.~(\ref{Eq3-11}), would lead to the enhancement of about 2 and 1.2 for the two cases   with and without a first-order phase transition, respectively.    The value of about 0.38 for \pdt in the case of $g_V=G_S$ is consistent with that from a recent study~\cite{Sun:2020uoj} based on the AMPT model with a smooth partonic to hadronic phase transition, which also predicts a constant value of around 0.4 for \pdt regardless of the collision energy.  A pure hadronic JAM transport model with light nuclei produced via a simple coalescence model based on some coalescence radii in phase space also predicts a  similar collision energy independent  constant value for \pdt~\cite{Liu:2019nii}. Compared to these baseline results from transport models, the enhancement in the yield ratio \pdt due to the first-order chiral phase transition in the present study is about 1.29, which is appreciable in comparison to the peak enhancement of 1.5 at $\sqrt{s_{NN}}\sim 20$ GeV in the preliminary data from the STAR Collaboration~\cite{Adam:2019wnb,Zhang:2019wun,Zhang:2020ewj}. Our findings thus justify the suggestion in Refs.~\cite{Sun:2017xrx,Sun:2018jhg} that the non-monotonic behavior in the collision energy dependence of the ratio \pdt is a good signature for the QCD phase transitions.

A possible enhanced production of light nuclei in the  presence of a first-order phase transition has also been found in a previous study based on a fluid dynamic model using the equation of state from the Polyakov Quark Meson model that has a first-order phase transition~\cite{Steinheimer:2013xxa}.  In this study,  the much heavier $A=5$ and $A=8$ nuclei were considered, which are, however, much harder to measure in high energy heavy ion collisions due to their very small numbers.  Besides, the hadronic effects are not considered in Ref.~\cite{Steinheimer:2013xxa}.

\section{conclusions and outlook}

To study how density fluctuations, which results from a first-order phase transition between the produced partonic matter to the hadronic matter in relativistic heavy ion collisions, evolve under subsequent hadronic scatterings,  we have extended the AMPT model by replacing its partonic cascade stage with a partonic transport model based on  the 3-flavor NJL model.  We have considered two equations of state  in the NJL model  for the partonic phase, with one having a first-order  chiral  transition and a critical point ($g_V=0$) and the other having only a crossover  chiral  transition ($g_V=G_S$).   With an initially thermalized partonic  fireball,  we have found that  for the case with a first-order   chiral phase transition, large density fluctuations  can develop in the partonic phase as the evolution trajectory of the system passes through the spinodal region of the QCD phase diagram during its expansion. These density fluctuations are found to carry over    entirely  to the initial hadronic matter after hadronization, and largely survive after hadronic scatterings. Using the nucleon coalescence model for nuclei production from nucleons, which is known to allow the inclusion of the effect of nucleon density fluctuations, we have calculated the yields of deuterons and tritons and found an enhancement in the yield ratio of $ N_\text{t}N_\text{p}/ N_\text{d}^2$ compared to that obtained   from  the case with a smooth crossover transition.  The present study thus constitutes an important step towards the understanding of the measured non-monotonic behavior in the collision energy dependence of the ratio \pdt and the determination of  the QCD phase structure through the production of light nuclei in heavy-ion collisions.

Since the  test particle method used in the present study suppresses the event-by-event fluctuations, which is  expected to be  significantly enhanced in the critical region, the present approach thus needs to be modified to  address the critical phenomena associated with the critical point in the QCD phase diagram.    As suggested in Refs.~\cite{Shuryak:2019ikv,Shuryak:2020},  the increase in the range of the nucleon-nucleon interaction in the vicinity of a critical point can   further enhance the production of light nuclei  in relativistic heavy ion collisions. It will be of great interest to extend the present study based on the mean-field approximation  to  include the  long-range correlation between nucleons near the critical point.  Also, to fully understand the collision energy dependence of the experimental results on light nuclei production, more realistic equations of state, such as that from the PNJL model~\cite{PhysRevD.77.114028,Li:2015pbv} or the extended NJL model with a scalar-vector intreaction~\cite{Sun:2020bbn}, and initial conditions for heavy ion collisions are needed.  Furthermore, a more realistic modelling of hadronization than that currently used in AMPT is also important. These improvements  will allow us to quantitatively understand the non-monotonic behavior of the yield ratio \pdt in the experimental data  and to locate the position of the critical point in the QCD phase diagram.

\begin{acknowledgements}
One of the authors  K.J.S. thanks Jorgan Randrup and Xiao-Feng Luo for helpful discussions. This work was supported in part by the US Department of Energy under Contract No.DE-SC0015266, the Welch Foundation under Grant No. A-1358, and  the National Natural Science Foundation of China under Grant No. 11922514 and  No. 11625521.
\end{acknowledgements}

\appendix
\section{Density fluctuations and light nuclei production in  the  thermal model}

In the conventional thermal model  for particle production   in relativistic heavy ion collisions, the produced matter  is assumed to be in global thermal and chemical equilibrium and to have a uniform density distribution. The effect of density fluctuations can be included in this model by assuming that the   produced matter  is in local thermal and chemical equilibrium  with a space dependent temperature $T({\bf x})$ and chemical potential $\mu({\bf x})$. For the simplified case of a constant temperature, the nucleon number density is  given by
\begin{eqnarray}
\rho_{n,p}({\bf x})&=& \frac{2}{(2\pi)^3}4\pi Tm^2K_2\left(\frac{m}{T}\right)\text{e}^{\frac{\mu_{n,p}({\bf x})}{T}}, \label{appeq1}
\end{eqnarray}
where $K_2$ is the  modified Bessel function of second kind and $\mu_{n,p}({\bf x})$ denotes the space dependent chemical potentials for neutron and proton.  Assuming that deuterons and tritons are in local thermal and chemical equilibriums with the nucleons, their number densities in this model are then
\begin{eqnarray}
\rho_{d}({\bf x})= \frac{3}{(2\pi)^3}4\pi T(2m)^2K_2\left(\frac{2m}{T}\right)\text{e}^{\frac{\mu_{n}({\bf x})+\mu_{p}({\bf x})}{T}}, \notag \\
\rho_{t}({\bf x})= \frac{2}{(2\pi)^3}4\pi T(3m)^2K_2\left(\frac{3m}{T}\right)\text{e}^{\frac{2\mu_{n}({\bf x})+\mu_{p}({\bf x})}{T}}.  \label{appeq2}
\end{eqnarray}
The yield ratio $\mathcal{O}_{\text{p-d-t}}= N_\text{t}N_\text{p}/N_\text{d}^2 $ can be calculated in the non-relativistic approximation as
\begin{eqnarray}
\mathcal{O}_{\text{p-d-t}} &=& \frac{K_2(\frac{m}{T})K_2(\frac{3m}{T})}{4(K_2(\frac{2m}{T}))^2} \frac{\int \text{d}^3{\bf{x}}\rho_{p} \int \text{d}^3{\bf x}\rho_n^2\rho_p}{[\int \text{d}^3{\bf x}\rho_n\rho_p]^2}  \notag \\
&\approx& \frac{1}{2\sqrt{3}}\frac{\int \text{d}^3{\bf{x}}\rho_{p} \int \text{d}^3{\bf x}\rho_n^2\rho_p}{[\int \text{d}^3{\bf x}\rho_n\rho_p]^2}\notag\\
&\approx &\frac{1}{2\sqrt{3}}  \frac{1+2C_\text{np}+\Delta\rho_n}{(1+C_\text{np})^2}.\label{appeq3}
\end{eqnarray}
which is identical to  Eq.~(\ref{Eq3-10}) obtained from the nucleon coalescence model. Density fluctuations thus  affect the yield ratio $N_\text{t}N_\text{p}/N_\text{d}^2 $ similarly in  both the coalescence and the thermal model.

The above derivation is based on the assumption that the local chemical equilibrium among protons, deuterons, and tritons is maintained from chemical freeze out until kinetic freeze out  as a result of the large  production and dissociation cross sections of deuterons and tritons during the hadronic evolution~\cite{Oliinychenko:2018ugs}.  If one assumes instead that the yields of deuterons and tritons  produced at the chemical freeze out of identified hadrons  remain  unchanged during hadronic evolution as assumed in usual statistical hadronization model, there is an additional factor in Eq.~(\ref{appeq3}) from the contribution of resonance decays to protons ~\cite{Oliinychenko:2020ply}.

\bibliography{myref}
\end{document}